\documentclass[a4paper,aps,prd,showpacs,showkeys,twocolumn,10pt]{revtex4-1}
\usepackage{graphicx,slashed,amssymb}
\usepackage{amsmath}
\def\be{\begin{eqnarray}}
\def\ee{\end{eqnarray}}
\def\nn{\nonumber\\}
\def\sumint{\sum\!\!\!\!\!\!\!\!\int_n}
\def\del{\partial}
\newcommand*{\mycdot}{\kern-.15em\cdot\kern-.15em}

\usepackage[colorlinks=true,linktocpage=true,linkcolor=blue,citecolor=blue]{hyperref}

\begin{document}
\title{Finite temperature QCD four-point function in the presence of a weak
    magnetic field within the hard thermal loop approximation}
\author{Najmul Haque}
\email{Najmul.Haque@theo.physik.uni-giessen.de}
\affiliation{Institut f\"ur Theoretische Physik, Justus-Liebig--Universit\"at Giessen, 35392 Giessen, Germany}

\begin{abstract}
We compute the weak magnetic field correction to the four-point quark-gluon vertex at finite temperature and chemical potential within the hard thermal loop approximation. The resulting four-point quark-gluon vertex satisfies the Ward identity with the three-point quark-gluon vertex.
\end{abstract}
\maketitle

\section{Introduction}
It is well known that quarks and gluons are confined within hadrons at low temperature and density where chiral symmetry is spontaneously broken. At high temperature and/or high density, chiral symmetry is restored, the hadronic states are believed to make a phase transition to a deconfined state of quarks and gluons, and a quasifree state of quarks and gluons, known as quark-gluon plasma, is produced. Such high-temperature and/or high-density situations are generated in ultrarelativistic heavy-ion collisions experiments at the Relativistic Heavy Ion Collider~\cite{bnlrhic} and Large Hadron Collider~\cite{ALICE} and are also expected to be generated in future experiment at the Facility for Antiproton and Ion Research~\cite{Friman:2011zz}. In such heavy-ion collisions, a very strong anisotropic magnetic field is generated~\cite{Fukushima:2008xe,Skokov:2009qp,Fukushima:2010vw,Voronyuk:2011jd,Deng:2012pc,Mueller:2014tea} due to the relative motion of the ions in the direction perpendicular to the reaction plane.

During the last few years, the properties of hot and dense nuclear matter in the presence of magnetic field have been the subject of intensive research. Several novel properties of hot and dense nuclear matter in the presence of the background magnetic field have been studied over the years, namely, the chiral magnetic effect~\cite{Kharzeev:2007jp,Fukushima:2008xe,Kharzeev:2009fn}; chiral- and color-symmetry broken/restoration phase~\cite{Gusynin:1997kj,Lee:1997zj,Chatterjee:2011ry,Chatterjee:2014csa}; magnetic catalysis~\cite{Alexandre:2000yf,Fayazbakhsh:2010gc,Mueller:2015fka} and inverse magnetic catalysis~\cite{Ayala:2014iba,Ayala:2014gwa,Farias:2014eca,Mueller:2015fka,Ayala:2015bgv}; bulk properties of Fermi gas~\cite{Strickland:2012vu}; phase structure of QCD~\cite{Bornyakov:2013eya,Bali:2011qj,Andersen:2014xxa,Fayazbakhsh:2010bh,Fayazbakhsh:2010gc}; various properties of mesons such as the decay constant, thermal mass, and dispersion relations~\cite{Andersen:2012zc,Fayazbakhsh:2012vr,Fayazbakhsh:2013cha,Adhya:2016ydf,Ghosh:2016evc}; soft photon production from conformal anomaly in heavy-ion collisions~\cite{Basar:2012bp,Ayala:2016lvs}; modification of QED dispersion properties~\cite{Sadooghi:2015hha}; electromagnetic radiation~\cite{Tuchin:2012mf}; dilepton production~\cite{Bandyopadhyay:2016fyd,Bandyopadhyay:2017raf,Tuchin:2013bda,Tuchin:2013ie,Sadooghi:2016jyf,Mamo:2013efa}; transport properties~\cite{Kadam:2014xka,Hattori:2016cnt}; and properties of quarkonia~\cite{Machado:2013rta,Alford:2013jva,Bonati:2015dka,Hasan:2017fmf}.

Another phenomenological important quantity for studying the hot and dense nuclear matter is the QCD equation of state (EoS). A lot of work has been done to study the QCD EoS in the absence of any background magnetic field using both the nonperturbative method like Lattice QCD (LQCD)~\cite{Allton:2005gk,Borsanyi:2012cr,Bazavov:2014pvz,Bellwied:2015lba} and also the hard thermal loop (HTL) resummed perturbation theory~\cite{Andersen:1999va,Andersen:1999sf,Andersen:1999fw,Andersen:2002ey,Andersen:2003zk,Chakraborty:2001kx,Chakraborty:2002yt,Chakraborty:2003uw,Andersen:2009tc,Andersen:2010ct,Andersen:2010wu,Andersen:2011sf,Haque:2010rb,Haque:2011vt,Haque:2011iz,Haque:2012my,Haque:2013qta,Haque:2013sja,Haque:2014rua,Andersen:2015eoa}. Recently, the QCD EoS in the presence of magnetic field has been studied using LQCD~\cite{Bali:2014kia,Bonati:2013vba}, the hadron resonance gas model~\cite{Endrodi:2013cs}, the field correlator method~\cite{Castorina:2015ava}, the Polyakov linear-sigma model~\cite{Tawfik:2016gye}, and also the perturbation theory in two-loop order~\cite{Blaizot:2012sd}. Very recently, in Ref.~\cite{Bandyopadhyay:2017cle}, QCD pressure has been computed in one-loop order in the presence of weak magnetic field applying HTL perturbation theory. As next-to-next-leading order (NNLO)  hard thermal loop
perturbation theory (HTLpt) thermodynamic results~\cite{Haque:2014rua} at vanishing magnetic field matches well with LQCD data, it will be interesting to extend the NNLO HTLpt calculations at nonvanishing magnetic field. To compute that, one needs various $N$-point functions at finite temperature and chemical potential in the presence of finite magnetic field. Recently, the quark propagator and quark-gluon vertex have been computed in the presence of a weak magnetic field in Ref.~\cite{Ayala:2014uua} within the HTL approximation. In this article, we calculate the only remaining $N$-point function, viz., the four-point quark-gluon vertex, in the presence of magnetic field, which would be essential for computing NNLO HTLpt thermodynamics. 

The paper is organized as follows. In Sec.~\ref{sec:23point}, we revisit the calculations for quark self-energy and the three-point quark-gluon vertex that have been done earlier in Ref.~\cite{Ayala:2014uua}. In Sec.~\ref{sec:4point}, we discuss the detailed calculations for the four-point quark-gluon vertex in the absence as well as in the presence of weak magnetic field. In Sec.~\ref{sec:conclusion}, we conclude with our results. 

\section{Two- and three-point functions in QCD}
\label{sec:23point}
The two- and three-point functions in QCD at weak magnetic field have been done earlier in Ref.~\cite{Ayala:2014uua} with vanishing chemical potential. The authors in Ref.~\cite{Ayala:2014uua} have explicitly shown (before doing the Matsubara frequency sum and three-momentum integration) that the three-point vertex satisfies the Ward identity with the quark self-energy. But in the final expression of the quark self-energy, the authors have missed a factor of $(-2)$.

After correcting that missing $(-2)$ factor and also considering the finite chemical potential of the system, the weak magnetic field correction to the quark self-energy at finite temperature and chemical potential can be written as
\be
\delta\Sigma (P) &=& - 4i\gamma_5g^2C_FM^2(T,\mu,m,qB)\nn
&\times& \int\frac{d\Omega}{4\pi} \frac{\left[ (\hat{K}\mycdot b)\slashed{u} - 
(\hat{K}\mycdot u)\slashed{b}\right]}
{(P\mycdot\hat{K})},
\label{selfenergy}
\ee
where $\hat{K}=(-i,{\bf\hat k})$, $C_F=(N_c^2-1)/2N_c$ with $N_c$ number of 
colors and the magnetic mass
\be
M^2(T,\mu,m,qB)=\frac{qB}{16\pi^2}\left[ -\frac{1}{4}\aleph(z)-\frac{\pi 
T}{2m}-\frac{\gamma_E}{2}\right].
\label{MB}
\ee
We use the gamma matrices relation to get Eq.~(\ref{selfenergy}),
\be
\gamma_1\gamma_2\slashed{K}_\shortparallel=\gamma_5\big[(K\mycdot 
b)\slashed{u}-(K\mycdot u)\slashed{b}\big],
\ee
with $u_\mu=(1,0,0,0)$ and $b_\mu=(0,0,0,1)$.

The function $\aleph(z)$ appears in Eq.~(\ref{MB}) due to the inclusion of finite chemical potential. With $z=\left(1/2-i\mu/(2\pi T)\right)=\left( 1/2-i\hat\mu\right)$, $\aleph(z)$ can be written as
\be
\aleph(z)= \Psi(z)+\Psi(z^*),
\ee
where $\Psi(z)= \Gamma'(z)/\Gamma(z)$ represents digamma function.

Also, the weak magnetic field correction of the three-point quark-gluon vertex can be written from Ref.~\cite{Ayala:2014uua} as
\be
  \delta\Gamma_\mu(P_1,P_2)&=&4i\gamma_5g^2C_FM^2(T,\mu,m,qB)\nn
 &\times& 
  \int\frac{d\Omega}{4\pi}\frac{1}{(P_1\cdot\hat{K})(P_2\cdot\hat{K})}\nn
&\times&
  \bigg\{\left[(\hat{K}\mycdot b)u_\mu - (\hat{K}\mycdot u)b_\mu\right]\hat{\slashed{K}}\nn
&&+ \left[(\hat{K}\mycdot b)\slashed{u} - (\hat{K}\mycdot u)\slashed{b}\right]\hat{K}_\mu
  \bigg\}.
\label{Gmuexpl}
\ee
One can now easily check the Ward identity from Eqs.~(\ref{selfenergy}) and (\ref{Gmuexpl}) as
\be
(P_1-P_2)^\mu\!\!\!&&\!\!\!\delta\Gamma_\mu(P_1,P_2) = 
4i\gamma_5g^2C_FM^2(T,\mu,m,qB)\nn
 &\times& \int\frac{d\Omega}{4\pi}\frac{1}{(P_1\mycdot\hat{K})(P_2\mycdot\hat{K})}\nn
 &\times&\bigg\{\!\!\left[(\hat{K}\mycdot b)(P_1-P_2)\mycdot u - (\hat{K}\mycdot u)(P_1-P_2)\mycdot b\right]\hat{\slashed{K}}\nn
 &+&\left[ (\hat{K}\mycdot b)\slashed{u} - (\hat{K}\mycdot u)\slashed{b}\right]\left[P_1\mycdot\hat{K}- P_2\mycdot\hat{K}\right]
 \!\! \bigg\}.
 \label{WI_23}
\ee
The first term within the curly brackets in Eq.~(\ref{WI_23}) can be neglected~\cite{Ayala:2014uua} by virtue of the HTL approximation and the remaining term can be written as
\be
(P_1-P_2)^\mu\!\!\!&&\!\!\!\delta\Gamma_\mu(P_1,P_2) \simeq 
4i\gamma_5g^2C_FM^2(T,\mu,m,qB)\nn
 &\times& \int\frac{d\Omega}{4\pi}\frac{1}{(P_1\mycdot\hat{K})(P_2\mycdot\hat{K})}
   \nn
   &+&
   \left[ (\hat{K}\mycdot b)\slashed{u} - (\hat{K}\mycdot u)\slashed{b}\right]\left[P_1\mycdot\hat{K}- P_2\mycdot\hat{K}\right]
 \!\! \bigg\}\nn
 &=& \delta\Sigma (P_1) -\delta\Sigma (P_2).
 \ee
\section{Four-point functions in QCD}
\label{sec:4point}
In QCD, the HTL resummed quark-gluon four-point vertex function 
$\Gamma_{\mu\nu}$ is defined via
\be
\delta^{ab}\Gamma_{\mu\nu,ab}^{ij} = g^2C_F\delta_{ij}\Gamma_{\mu\nu}.
\label{Gmunu_def}
\ee
Now, the following diagrams will contribute to the HTL resummed quark-gluon four-point vertex function.
\begin{figure}[tbh]
 \includegraphics[width=8.5cm,height=7cm]{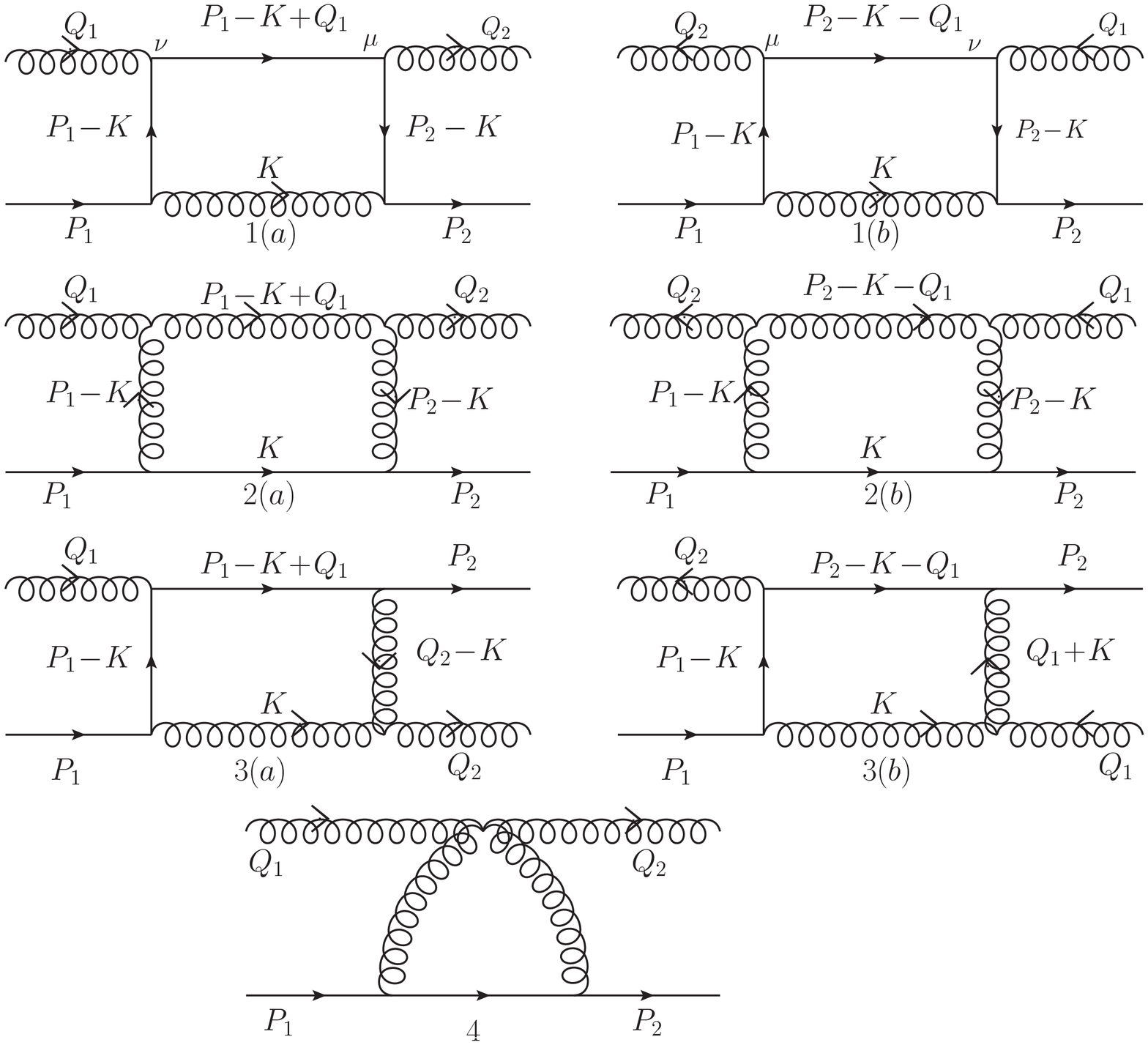}
 \caption{Feynman diagrams for the four-point quark gluon vertex.}
 \label{feyn_dia}
\end{figure}
\subsection{At vanishing magnetic field}
In this subsection, we present the calculations for the four-point quark-gluon vertex function in the absence of any magnetic field to generalize the QED result presented in Ref.~\cite{Bellac:2011kqa}. The frequency sums that appear in this subsection will also be useful to calculate the frequency sums in the presence of magnetic field. Now, the color-traced four-point function for the diagram $1(a)$ can be written as
\be
\delta^{ab}\Gamma_{\mu\nu,ab}^{ij,\ 1(a)}&=& g^4C_F^2 
\delta_{ij}T\sumint\frac{d^3k}{(2\pi)^3}\gamma_\alpha
(\slashed{K}-\slashed{P}_2)\gamma_\mu\nn
&\times&(\slashed{K}-\slashed{P}_1-\slashed{Q}_1)\gamma_\nu(\slashed{K} - 
\slashed{P}_1)\gamma_\alpha\Delta(K)\nn
&\times&\!\!
\tilde{\Delta}(K\!-\!P_2)\tilde{\Delta}(K\!-\!P_1\!-\!Q_1\!)\tilde{\Delta}
(K\!-\!P_1\!),\ 
\label{gamma_a_B0}
\ee
where
\be
\tilde{\Delta}(K)&=&\frac{1}{((2n+1)\pi T-i\mu)^2 +k^2 +m^2},
\label{Delta_f}\\
\Delta(K)&=&\frac{1}{((2n+1)\pi T)^2 +k^2}.
\label{Delta_b}
\ee
In the HTL approximation we neglect the quark mass $m$ at high temperature in the absence of any background magnetic field. However, it is important to keep a nonzero quark mass in the presence of magnetic field~\cite{Alexandre:2000jc,Ayala:2014uua}. Finite quark mass acts as an infrared scale at finite magnetic field, and in numerical computation, we can take the quark mass  as thermal quark mass as discussed in Ref.~\cite{Ayala:2014uua}.

Now, in the HTL approximation we can neglect external momenta with respect to the loop momenta~\cite{Braaten:1989mz}, and the four-point vertex function from diagram $1(a)$ of Fig.~\ref{feyn_dia} using Eqs.~(\ref{Gmunu_def}) and (\ref{gamma_a_B0}) can be written as
\be
\Gamma_{\mu\nu}^{1(a)}&\simeq& g^2C_F T\sumint\frac{d^3k}{(2\pi)^3}
\gamma_\alpha\slashed{K}\gamma_\mu \slashed{K}
\gamma_\nu \slashed{K}\gamma_\alpha\Delta(K)\nn
&\times&\tilde{\Delta}(K-P_1)\tilde{\Delta}(K-P_1-Q_1)\tilde{\Delta}(K-P_2),\nn
&=&g^2C_F T\sumint\frac{d^3k}{(2\pi)^3}\gamma_\alpha(4K_\mu K_\nu\slashed{K}+2K^2 
K_\mu\gamma_\nu \nn
&&- K^2\gamma_\mu\gamma_\nu\slashed{K})\gamma_\alpha\tilde{\Delta}(K-P_1)\nn
&\times&\tilde{\Delta}(K-P_1-Q_1)\tilde{\Delta}(K-P_2)\Delta(K)
\nn
&\simeq& 8g^2C_FT\sumint\frac{d^3k}{(2\pi)^3} K_\mu 
K_\nu\slashed{K}\tilde{\Delta}(K-P_1)\nn
&\times&\tilde{\Delta}(K-P_1-Q_1)\tilde{\Delta}(K-P_2) \Delta(K).
\label{gamma_1a_B0}
\ee
Similarly,
\be
\Gamma_{\mu\nu}^{2(a)} &\simeq& 8g^2C_A T\sumint\frac{d^3k}{(2\pi)^3} K_\mu K_\nu \slashed{K}\Delta(K\!-\!P_1)\nn
&\times&\Delta(K-P_1-Q_1)\Delta(K-P_2)\tilde{\Delta}(K).
\label{gamma_2a_B0}
\ee
and
\be
\Gamma_{\mu\nu}^{3(a)} &\simeq& - 4g^2C_AT\sumint\frac{d^3k}{(2\pi)^3} K_\mu K_\nu \slashed{K}\tilde{\Delta}(K\!-\!P_1)\nn
&\times&\tilde{\Delta}(K-P_1-Q_1)\Delta(K-Q_2)\Delta(K).
\label{gamma_3a_B0}
\ee
One does not need to compute the diagrams $1(b), 2(b),$ and $3(b)$ explicitly. One can obtain the contributions from these diagrams by replacing $P_1+Q_1\rightarrow P_2-Q_1$ from the corresponding first diagram of each set. The remaining diagram (diagram 4) does not contribute anything in the HTL approximation to the four-point vertex.

To calculate the frequency sum of Eqs.~(\ref{gamma_1a_B0}) -(\ref{gamma_3a_B0}), we can define the following tensors of which the components need to be computed:
\be
\mathcal{T}^{1(a)}_{\mu\nu\beta} &=& 
T\sumint\frac{d^3k}{\left(2\pi\right)^3} K_\mu K_\nu K_\beta
\tilde{\Delta}(P_1-K)\nn
&\times& \Delta(K)\tilde{\Delta}(P_2-K)\tilde{\Delta}(P_1-K+Q_1),
\label{tensor_1a_B0}
\\
\mathcal{T}^{2(a)}_{\mu\nu\beta} &=& 
T\sumint\frac{d^3k}{\left(2\pi\right)^3} K_\mu K_\nu K_\beta
\Delta(P_1-K)\nn
&\times& \tilde{\Delta}(K)\Delta(P_2-K)\Delta(P_1-K+Q_1),
\label{tensor_2a_B0}
\\
\mathcal{T}^{3(a)}_{\mu\nu\beta} &=& 
T\sumint\frac{d^3k}{\left(2\pi\right)^3} K_\mu K_\nu K_\beta
\tilde{\Delta}(P_1-K)\nn
&\times& \Delta(K)\Delta(Q_2-K)\tilde{\Delta}(P_1-K+Q_1).
\label{tensor_3a_B0}
\ee
Now, the following Matsubara sums corresponding to Eqs.~(\ref{tensor_1a_B0})-(\ref{tensor_3a_B0}) are to be evaluated.
\be
X_{i}^{1(a)}&=&T\sum\limits_{n}\omega_n^i\Delta(K)\tilde{\Delta}(P_1-K)\nn
&\times&\tilde{\Delta}(P_2-K)\tilde{\Delta}(P_1-K+Q_1),
\label{X_1a}
\\
X_{i}^{2(a)}&=&T\sum\limits_{n}\omega_n^i\tilde{\Delta}(K)\Delta(P_1-K)\nn
&\times&\Delta(P_2-K)\Delta(P_1-K+Q_1),
\label{X_2a}
\\
X_{i}^{3(a)}&=&T\sum\limits_{n}\omega_n^i\Delta(K)\tilde{\Delta}(P_1-K)\nn
&\times&\Delta(Q_2-K)\tilde{\Delta}(P_1-K+Q_1).
\label{X_3a}
\ee
\begin{widetext}
For $i=0$, Eq.~(\ref{X_1a}) reduces to
\be
X_0^{1(a)}
&=&T\sum\limits_{n}\Delta(K)\tilde{\Delta}(P_1-K)\tilde{\Delta}(P_2-K)\tilde{
\Delta}(P_1-K+Q_1)\nn
&=&\frac{1}{16EE_1E_2E_3}\sum\limits_{s,s_1,s_2,s_3}
\frac{ss_1s_2s_3}{i(\omega_{1\mu}-\omega_{2\mu}+\omega_3)-s_3E_3+s_2E_2}\nn
&\times&\Bigg[\frac{1}{i\omega_3+s_1E_1-s_3E_3}\Bigg\{\frac{
1+n_B(sE)-n_F^+(s_1E_1)}{i\omega_{1\mu}
 -sE-s_1E_1}
-\frac{1+n_B(sE)-n_F^+(s_3E_3)}{i(\omega_{1\mu}+\omega_3)-sE-s_3E_3}\!\Bigg\}\nn
&&+\frac{1}{i(\omega_{1\mu}
-\omega_{2\mu})-s_1E_1 + s_2E_2} 
\Bigg\{\!\!\frac{1+n_B(sE)-n_F^+(s_1E_1)}{i\omega_{1\mu}
 -sE-s_1E_1} - 
\frac{1+n_B(sE)-n_F^+(s_2E_2)}{i\omega_{2\mu}-sE-s_2E_2}\!\Bigg\}\!\Bigg],\hspace
{.6cm}
\label{X01a_sum}
 \ee
 \end{widetext}
where
\be
P_1&=&(p_{10},{\bf p_1})=(-\omega_1+i\mu,{\bf p_1})=(-\omega_{1\mu},{\bf 
p_1}),\nn
P_2&=&(-\omega_{2\mu},{\bf p_2}),
Q_1=(-\omega_{3},{\bf q_1}),\ \ 
\ee
and also
\be
E&=&k;\ E_1=|{\bf k-p_1}|;\  E_2=|{\bf k-p_2}|;\nn
E_3&=&|{\bf k-p_1-q_1}|.
\ee
The functions $n_B(x)$ and $n_F^\pm(x)$ in Eq.~(\ref{X01a_sum}) represent 
Bose-Einstein and Fermi-Dirac distribution
functions and can be expressed as
\be
n_B(x)=\frac{1}{e^{x/T}-1},\\
n_F^\pm(x)=\frac{1}{e^{(x\mp\mu)/T}+1}.
\ee
The leading-order term only appears for $-s=s_1=s_2=s_3$. So, the leading-order contribution from Eq.~(\ref{X01a_sum}) at the high-temperature limit can be written as
\begin{widetext}
\be
X_0^{1(a)}&\simeq&-\frac{1}{16EE_1E_2E_3}\frac{1}{i(\omega_{1\mu}-\omega_{2\mu}
+\omega_3)+E_3-E_2}\Bigg[
\frac{1}{i\omega_3-E_1+E_3}\Bigg\{\frac{n_B(E)+n_F^-(E_1)}{i\omega_{1\mu}
 -E+E_1}\nn
 &-& \frac{n_B(E)+n_F^-(E_3)}{i(\omega_{1\mu}+\omega_3)-E+E_3}\Bigg\}
 +\frac{1}{i(\omega_{1\mu}-\omega_{2\mu})+E_1 - E_2} 
\Bigg\{\frac{n_B(E)+n_F^-(E_1)}{i\omega_{1\mu}
 -E+E_1} - \frac{n_B(E)+n_F^-(E_2)}{i\omega_{2\mu}-E+E_2}\Bigg\}\Bigg]\nn
 &+& 
\frac{1}{16EE_1E_2E_3}\frac{1}{i(\omega_{1\mu}-\omega_{2\mu}+\omega_3)-E_3+E_2}
\Bigg[
\frac{1}{i\omega_3+E_1-E_3}\Bigg\{\frac{n_B(E)+n_F^+(E_1)}{i\omega_{1\mu}
 +E-E_1}\nn 
&-&\frac{n_B(E)+n_F^+(E_3)}{i(\omega_{1\mu}+\omega_3)+E-E_3}\Bigg\}+\frac{1}{
i(\omega_{1\mu}-\omega_{2\mu})-E_1 + E_2}
 \Bigg\{\frac{n_B(E)+n_F^+(E_1)}{i\omega_{1\mu}
 +E-E_1}-\frac{n_B(E)+n_F^+(E_2)}{i\omega_{2\mu}+E-E_2}\Bigg\}\Bigg].
\ee
In the HTL approximation we may use
\be
E&=&k;\ \ E_1\simeq k-{\bf p_1\mycdot k};\ \ E_2\simeq k-{\bf p_2\mycdot k};\nn
E_3&\simeq&k-({\bf p_1+q_1)}\mycdot{\bf k},
\ee
and also
$ n_F^\pm(E_i)\simeq n_F^\pm(k)$ to simplify $X_0^{1(a)}$ as
\be
X_0^{1(a)}&\simeq&\frac{n_B(k)+n_F^-(k)}{16k^4}
\Bigg\{\frac{1}{(i\omega_{1\mu}
 -E+E_1)(i\omega_{2\mu}-E+E_2)}\frac{1}{(i(\omega_{1\mu}+\omega_3)-E+E_3)}\Bigg\}\nn
 &-& 
\frac{n_B(k)+n_F^+(k)}{16k^4}\Bigg\{\frac{1}{(i\omega_{1\mu}
 +E-E_1)(i\omega_{2\mu}+E-E_2)}\frac{1}{(i(\omega_{1\mu}+\omega_3)+E-E_3)}\Bigg\}\nn
 &=& \frac{1}{16k^4}
\Bigg[\frac{n_B(k)+n_F^-(k)}{(P_1\mycdot\hat{K}')(P_2\mycdot\hat{K}')[
(P_1+Q_1)\mycdot\hat{K}']} - 
\frac{n_B(k)+n_F^+(k)}{(P_1\mycdot\hat{K})(P_2\mycdot\hat{K})[(P_1 + 
Q_1)\mycdot\hat{K}]}\Bigg],
\label{X0_1a_f}
\ee
\end{widetext}
with $\hat{K}=(-i,\hat{\bf k})$,\ \ $\hat{K}'=(-i,-\hat{\bf k})$.
Similarly for the diagram $1(b)$, we can write
\be
X_0^{1(b)}&=&\frac{1}{16k^4}
\Bigg[\frac{n_B(k)+n_F^-(k)}{(P_1\mycdot\hat{K}')(P_2\mycdot\hat{K}
')[(P_2-Q_1)\mycdot\hat{K}']}\nn
 && - 
\frac{n_B(k)+n_F^+(k)}{(P_1\mycdot\hat{K})(P_2\mycdot\hat{K})[
(P_2-Q_1)\mycdot\hat{K}]}\Bigg].
\label{X0_1b_f}
\ee
The other frequency sums ($i=1,2,3$) of Eq.~(\ref{X_1a}) can be calculated using Eq.~(\ref{X0_1a_f}). $X_1^{1(a)}$ can be obtained by replacing $\omega_n\rightarrow -isE$ before doing the summations in Eq.~(\ref{X01a_sum}) and can be written as
\be
X_1^{1(a)}&=&-\frac{i}{16k^3}
\Bigg[\frac{n_B(k)+n_F^-(k)}{(P_1\mycdot\hat{K}')(P_2\mycdot\hat{K}')[
(P_1+Q_1)\mycdot\hat{K}']}\nn
&& +
\frac{n_B(k)+n_F^+(k)}{(P_1\mycdot\hat{K})(P_2\mycdot\hat{K})[(P_1+Q_1)\mycdot\hat{K}]}
\Bigg].
\label{X1_1a_f}
\ee
$X_2^{1(a)}$ can be obtained using the property that any $K^2$ term in the numerator gives a nonleading contribution and hence
\be
X_2^{1(a)}&\simeq&-k^2X_0^{1(a)}\nn
&=&-\frac{1}{16k^2}\Bigg[\frac{n_B(k)+n_F^-(k)}{(P_1\mycdot\hat{K}')(P_2\mycdot\hat{K}')[(P_1+Q_1)\mycdot\hat{K}']}\nn
&& -\frac{n_B(k)+n_F^+(k)}{(P_1\mycdot\hat{K})(P_2\mycdot\hat{K})[(P_1+Q_1)\mycdot\hat{K}]}\Bigg].
\label{X2_1a_f}
\ee
Similarly,
\be
X_3^{1(a)}&\simeq&-k^2X_1^{1(a)}\nn
&=&\frac{i}{16k}
\Bigg[\frac{n_B(k)+n_F^-(k)}{(P_1\mycdot\hat{K}')(P_2\mycdot\hat{K}')[(P_1+Q_1)\mycdot\hat{K}']}\nn
 &&+
\frac{n_B(k)+n_F^+(k)}{(P_1\mycdot\hat{K})(P_2\mycdot\hat{K})[(P_1+Q_1)\mycdot\hat{K}]}\Bigg].
\label{X3_1a_f}
\ee
Now, we calculate the next frequency sum that is given in Eq.~(\ref{X_2a}). As all the bosonic and the fermionic propagators get exchanged to get Eq.~(\ref{X_2a}) from Eq.~(\ref{X_1a}), we can get the expression for $X_{i}^{2(a)}$ by replacing $n_B(x)\leftrightarrow - n_F^+(x)$ in Eq.~(\ref{X01a_sum}) and hence
\begin{widetext}
\be
X_0^{2(a)}
&=&T\sum\limits_{n}\tilde{\Delta}
(K)\Delta(P_1-K)\Delta(P_2-K)\Delta(P_1-K+Q_1)\nn
&=&\frac{1}{16EE_1E_2E_3}\sum\limits_{s,s_1,s_2,s_3}
\frac{1}{i(\omega_{1\mu}-\omega_{2\mu}+\omega_3)-s_3E_3+s_2E_2}\nn
&\times&\Bigg[\frac{1}{i\omega_3+s_1E_1-s_3E_3}\Bigg\{\frac{
1-n_F^+(sE)+n_B(s_1E_1)}{i\omega_{1\mu}
 -sE-s_1E_1}
-\frac{1-n_F^+(sE)+n_B(s_3E_3)}{i(\omega_{1\mu}+\omega_3)-sE-s_3E_3}\!\Bigg\}\nn
&&+\frac{1}{i(\omega_{1\mu}
-\omega_{2\mu})-s_1E_1 + s_2E_2} 
\Bigg\{\!\!\frac{1-n_F^+(sE)+n_B(s_1E_1)}{i\omega_{1\mu}
 -sE-s_1E_1} - 
\frac{1-n_F^+(sE)+n_B(s_2E_2)}{i\omega_{2\mu}-sE-s_2E_2}\!\Bigg\}\!\Bigg].\hspace
{.6cm}
\label{X02a_sum}
 \ee
By keeping only the terms that will contribute to the leading order and also applying the HTL approximation, Eq.~(\ref{X02a_sum}) reduces to 
\be
X_0^{2(a)} &=& -\frac{1}{16k^4}
\Bigg[\frac{n_B(k)+n_F^-(k)}{(P_1\mycdot\hat{K}')(P_2\mycdot\hat{K}')[
(P_1+Q_1)\mycdot\hat{K}']}
- 
\frac{n_B(k)+n_F^+(k)}{(P_1\mycdot\hat{K})(P_2\mycdot\hat{K})[(P_1 + 
Q_1)\mycdot\hat{K}]}\Bigg]
=-X_0^{1(a)}.
\label{X_2a_f}
\ee
Similarly, for the diagram $2(b)$, we can write

\be
X_0^{2(b)} &=& -\frac{1}{16k^4}
\Bigg[\frac{n_B(k)+n_F^-(k)}{(P_1\mycdot\hat{K}')(P_2\mycdot\hat{K}')[
(P_2-Q_1)\mycdot\hat{K}']} 
- 
\frac{n_B(k)+n_F^+(k)}{(P_1\mycdot\hat{K})(P_2\mycdot\hat{K})[(P_2 - 
Q_1)\mycdot\hat{K}]}\Bigg]
=-X_0^{1(b)}.
\label{X_2b_f}
\ee
One can explicitly compute all the summations in Eq.~(\ref{X_2a}) for all values of $i$, and it can be shown that
\be
X_i^{1(a)}=-X_i^{2(a)}.
\label{X1a2a}
\ee
Similarly, we can calculate the frequency sum in Eq.~(\ref{X_2a}) as
\be
X_0^{3(a)}
&=&T\sum\limits_{n}\Delta(K)\tilde{\Delta}(P_1-K)\Delta(Q_2-K)\tilde{\Delta}
(P_1-K+Q_1)\nn
&=&-\frac{1}{16EE_1E_3E_4}\sum\limits_{s,s_1,s_3,s_4}
\frac{1}{i\omega_3+s_1E_1-s_3E_3}\frac{1}{i\omega_4-sE+s_4E_4}
\Bigg[\frac{1+n_B(sE)-n_F^+(s_3E_3)}{i(\omega_{1\mu}+\omega_3)-sE-s_3E_3}\nn
&-&\frac{1+n_B(sE)- n_F^+(s_1E_1)}{i\omega_{1\mu}-sE-s_1E_1}
- \frac{1+n_B(s_4E_4)-n_F^+(s_3E_3)}{i(\omega_{1\mu}+\omega_3-\omega_4)
 -s_3E_3-s_4E_4} + 
\frac{1+n_B(s_4E_4)-n_F^+(s_1E_1)}{i(\omega_{1\mu}-\omega_4)-s_1E_1-s_4E_4}\Bigg
],
\label{X03a_sum}
 \ee
where
\be
Q_2=(-\omega_{4},{\bf q_2});\ E_4=|{\bf k-q_2}|.
\ee
The leading-order term only appears for $s=-s_1=-s_2=s_4$ and can be written from Eq.~(\ref{X03a_sum}) as
\be
X_0^{3(a)}&=&T\sum\limits_{n}\Delta(K)\tilde{\Delta}(P_1-K)\Delta(Q_2-K)\tilde
{\Delta}(P_1-K+Q_1)\nn
&=&-\frac{1}{16EE_1E_3E_4}
\frac{1}{i\omega_3-E_1+E_3}\frac{1}{i\omega_4-E+E_4}
\Bigg[\frac{n_B(E)+n_F^-(E_3)}{i(\omega_{1\mu}+\omega_3)-E+E_3}\nn
&-&\frac{n_B(E) + n_F^-(E_1)}{i\omega_{1\mu}-E+E_1}
- \frac{n_B(E_4)+n_F^-(E_3)}{i(\omega_{1\mu}+\omega_3-\omega_4)
 +E_2-E_4} + 
\frac{n_B(E_4)+n_F^-(E_1)}{i(\omega_{1\mu}-\omega_4)+E_1-E_4}\Bigg]\nn
&+&\frac{1}{16EE_1E_3E_4}
\frac{1}{i\omega_3+E_1-E_3}\frac{1}{i\omega_4+E-E_4}
\Bigg[\frac{n_B(E)+n_F^+(E_3)}{i(\omega_{1\mu}+\omega_3)+E-E_3}\nn
&-&\frac{n_B(E)+n_F^+(E_1)}{i\omega_{1\mu}+E-E_1}
- \frac{n_B(E_4)+n_F^+(E_3)}{i(\omega_{1\mu}+\omega_3-\omega_4)
 -E_2+E_4} + 
\frac{n_B(E_4)+n_F^+(E_1)}{i(\omega_{1\mu}-\omega_4)-E_1+E_4}\Bigg],\nn
 &\simeq&- \frac{1}{16k^4}
\Bigg[\Big\{n_B(k)+n_F^-(k)\Big\}\left(\frac{1}{P_1\mycdot\hat{K}'}+\frac{1}{
P_2\mycdot\hat{K}'}\right)\frac{1}{[(P_1+Q_1)\mycdot\hat{K}'][
(P_2-Q_1)\mycdot\hat{K}']}\nn
&&\hspace{.7cm}-\Big\{n_B(k)+n_F^+(k)\Big\}\left(\frac{1}{P_1\mycdot\hat{K}}
+\frac{1}{P_2\mycdot\hat{K}}\right)\frac{1}{[(P_1+Q_1)\mycdot\hat{K}][
(P_2-Q_1)\mycdot\hat{K}]} 
\Bigg].
\label{X_3a_f}
\ee
Similarly, for the diagram $3(b)$, we can write
\be
X_0^{3(b)}
 &\simeq&- \frac{1}{16k^4}
\Bigg[\Big\{n_B(k)+n_F^-(k)\Big\}\left(\frac{1}{P_1\mycdot\hat{K}'}+\frac{1}{
P_2\mycdot\hat{K}'}\right)\frac{1}{[(P_2-Q_1)\mycdot\hat{K}'][
(P_1+Q_1)\mycdot\hat{K}']}\nn
&&\hspace{1.05cm}-\Big\{n_B(k)+n_F^+(k)\Big\}\left(\frac{1}{P_1\mycdot\hat{K}}
+\frac{1}{P_2\mycdot\hat{K}}\right)\frac{1}{[(P_2-Q_1)\mycdot\hat{K}][
(P_1+Q_1)\mycdot\hat{K}]} 
\Bigg].
\label{X_3b_f}
\ee
\end{widetext}
Using Eqs.~(\ref{X_2a_f}),~(\ref{X_2b_f}),~(\ref{X_3a_f}) and~(\ref{X_3b_f}), 
we can write
\be
X_{0}^{2(a)}+X_{0}^{2(b)}=\frac{1}{2}\Big(X_{0}^{3(a)}+X_{0}^{3(b)}\Big).
\label{X_123a}
\ee

So, it is clear from Eqs.~(\ref{gamma_2a_B0}) and (\ref{gamma_3a_B0}) that the total contribution from the second sets of diagrams $[2(a),2(b)]$ cancel with the third sets of diagrams $[3(a),3(b)]$ if one uses the relation defined in Eq.~(\ref{X_123a}).

Now, using the Matsubara sums~(\ref{X0_1a_f}),~(\ref{X1_1a_f}),~(\ref{X2_1a_f}) and~(\ref{X3_1a_f}) we can rewrite the tensor~(\ref{tensor_1a_B0}) as
\be
\mathcal{T}^{1(a)}_{\mu\nu\beta} &=&-\frac{1}{32\pi^2}\int 
kdk\Big[2n_B(k)+n_F^+(k)+n_F^-(k)\Big]\nn
&\times&\int\frac{d\Omega}{4\pi} \frac{\hat{K}_\mu\hat{K}_\nu\hat{K}_\beta}
{(P_1\mycdot\hat{K})(P_2\mycdot\hat{K})[(P_1 + Q_1)\mycdot\hat{K}]}\nn
&=&-\frac{T^2}{64}(1+4\hat\mu^2)\nn
&\times&\int\frac{d\Omega}{4\pi} \frac{\hat{K}_\mu\hat{K}_\nu\hat{K}_\beta}
{(P_1\mycdot\hat{K})(P_2\mycdot\hat{K})[(P_1 + Q_1)\mycdot\hat{K}]}
\label{tensor_1a_B0_2}
\ee
Finally, the total HTL resummed four-point vertex function at vanishing magnetic field can be written as
\be
\Gamma_{\mu\nu}&=&8g^2C_F T\sumint\frac{d^3k}{(2\pi)^3}K_\mu K_\nu
\slashed{K}\tilde{\Delta}(K-P_1)\nn
&\times&\tilde{\Delta}(K-P_1-Q_1)\tilde{\Delta}(K-P_2)\Delta(K)\nn
&=&-m_q^2\int\frac{d\Omega}{4\pi}\left[\frac{1}{P_1\mycdot\hat{K}}+\frac{1}{
P_2\mycdot\hat{K}}\right]
\nn
&&\times\frac{\hat{K}_\mu\hat{K}_\nu\slashed{\hat{K}}}
{[(P_1 + Q_1)\mycdot\hat{K}][(P_2 - Q_1)\mycdot\hat{K}]}
\label{gamma_munu_B0}
\ee
with the square of quark thermal mass
\be
m_q^2=\frac{g^2T^2C_F}{8}\left(1+4\hat\mu^2\right).
\ee
Note that Eq.~(\ref{gamma_munu_B0}) exactly reproduces the QED four-point vertex in Ref.~\cite{Bellac:2011kqa} if we replace quark thermal mass with electron thermal mass. One can also easily show that the four-point function presented in Eq.~(\ref{gamma_munu_B0}) satisfies the Ward identity with the QCD three-point vertex in the absence of any magnetic field~\cite{Bellac:2011kqa}.
\subsection{At nonvanishing magnetic field}
Now, we calculate the magnetic field correction in the first order in the weak field approximation of the four-point vertex function. Using the magnetic field approximated quark propagator from Ref.~\cite{Ayala:2014uua}, we can write $\mathcal{O}(qB)$ term of the diagram $1(a)$ as
\be
\delta\Gamma^{1(a)}_{\mu\nu}&\simeq& -i 
g^2C_F(qB)T\sumint\frac{d^3k}{(2\pi)^3}\gamma_\alpha\nn
&\times&\Big[(\gamma_1\gamma_2
\slashed{K}_\shortparallel)\gamma_\nu \slashed{K}
\gamma_\mu \slashed{K}\tilde{\Delta}(K-P_1)\nn
&+& \slashed{K}\gamma_\nu(\gamma_1\gamma_2
\slashed{K}_\shortparallel)\gamma_\mu \slashed{K} \tilde{\Delta}(K-P_1-Q_1)\nn
&+& \slashed{K}\gamma_\mu \slashed{K}
\gamma_\nu(\gamma_1\gamma_2\slashed{K}_\shortparallel)\tilde{\Delta}(K-P_2)\Big]
\gamma_\alpha\Delta(K)\nn
&\times&\tilde{\Delta}(K-P_2)\tilde{\Delta}(K\!-\!P_1\!-\!Q_1)\tilde{\Delta}
(K-P_1).
\label{gamma_1a_B1}
\ee
Now, we can get the relation using the gamma matrices identity:
\be
\gamma_1\gamma_2\slashed{K}_\shortparallel=\gamma_5\big[(K\mycdot 
b)\slashed{u}-(K\mycdot u)\slashed{b}\big].
\ee
So, Eq.~(\ref{gamma_1a_B1}) becomes
\be
\delta\Gamma^{1(a)}_{\mu\nu}&\simeq&i g^2C_F(qB)\gamma_5 T\sumint\frac{d^3k}{(2\pi)^3}\gamma_\alpha \nn
&\times&
\Big[\big\{(K\mycdot b)\slashed{u}-(K\mycdot u)\slashed{b}\big\}\gamma_\nu 
\slashed{K}\gamma_\mu \slashed{K}\nn
&+& \slashed{K}\gamma_\nu\big\{(K\mycdot b)\slashed{u}-(K\mycdot 
u)\slashed{b}\big\}\gamma_\mu \slashed{K} \nn
&+& \slashed{K}\gamma_\nu \slashed{K} \gamma_\mu\big\{(K\mycdot 
b)\slashed{u}-(K\mycdot u)\slashed{b}\big\} \Big]
\gamma_\alpha\Delta(K)\nn
&\times&\tilde{\Delta}(K-P_2)\tilde{\Delta}(K-P_1-Q_1)\tilde{\Delta}^2(K-P_1),
\nn
&\simeq& 8i g^2C_F\gamma_5T\sumint\frac{d^3k}{(2\pi)^3} \Bigg[\Big\{(K\mycdot b)(K_\mu 
u_\nu + K_\nu u_\mu)\nn
&-& 
(K\mycdot u)(K_\mu b_\nu + K_\nu b_\mu)\Big\}\slashed{K} \nn
&+& K_\mu K_\nu\Big\{(K\mycdot b)\slashed{u}-(K\mycdot 
u)\slashed{b}\Big\}\Bigg]\Delta(K)\nn
&\times&\tilde{\Delta}(K-P_2)\tilde{\Delta}(K\!-\!P_1\! -\! 
Q_1)\tilde{\Delta}^2(K-P_1)
\label{gamma_1a_B}
\ee
Similarly
\be
\delta\Gamma^{2(a)}_{\mu\nu}&=& 8ig^2C_A(qB)\gamma_5T\sumint\frac{d^3k}{\left(2\pi\right)^3} \nn
&\times&\Bigg[\Big\{(K\mycdot b)(K_\mu u_\nu + K_\nu u_\mu)\nn
&&- (K\mycdot u)(K_\mu b_\nu + K_\nu b_\mu)\Big\}\slashed{K} \nn
&+& K_\mu 
K_\nu\Big\{(K\mycdot b)\slashed{u}-(K\mycdot u)\slashed{b}\Big\}\Bigg]\Delta(K\!-\!P_1)\nn
&\times&\tilde{\Delta}^2(K)\Delta(K-P_2)\Delta(K-P_1-Q_1)
\label{gamma_2a_B}
\ee
and
\be
\delta\Gamma^{3(a)}_{\mu\nu}&=&-4ig^2C_A(qB)\gamma_5T\sumint\frac{d^3k}{\left(2\pi\right)^3}\nn
&\times&\Bigg[\Big\{(K\mycdot b)(K_\mu u_\nu + K_\nu u_\mu)\nn
&& -(K\mycdot u)(K_\mu b_\nu + K_\nu b_\mu)\Big\}\slashed{K} \nn
&+& K_\mu K_\nu\Big\{(K\mycdot b)\slashed{u}-(K\mycdot 
u)\slashed{b}\Big\}\Bigg]\tilde{\Delta}^2(K\!-\!P_1)\nn
&\times&\Delta(K)\Delta(K-P_2)\Delta(K-P_1-Q_1)
\label{gamma_3a_B}
\ee
As in the case of zero magnetic field, it is also possible to show here that
\be
\delta\Gamma^{2(a)}_{\mu\nu}+\delta\Gamma^{2(b)}_{\mu\nu}=-\left(\delta\Gamma^{3(a)}_{\mu\nu}
+\delta\Gamma^{3(b)}_{\mu\nu}\right).
\ee
So, to calculate the weak magnetic field correction to the QCD four-point function, one needs to consider only $\delta\Gamma^{1(a)}$ and $\delta\Gamma^{1(b)}$. To calculate the frequency sum of $\delta\Gamma^{1(a)}$ as defined in Eq.~(\ref{gamma_1a_B}), we can define the following tensor of which the components need to be computed. 
\be
\delta\mathcal{T}^{1(a)}_{\mu\nu\beta} &=& 
T\sumint\frac{d^3k}{(2\pi)^3}K_\mu K_\nu K_\beta 
\tilde{\Delta}(P_2-K)\nn
&\times&\Delta(K) \tilde{\Delta}(P_1-K+Q_1)\tilde{\Delta}^2(P_1-K),
\label{tensor_def}
\ee
which in turn requires one to compute the frequency sums
\be
Y_{i}^{1(a)}&=&T\sum\limits_{n}\omega_n^i\Delta(K)\tilde{\Delta}^2(P_1-K)\nn
&\times&\tilde{\Delta}(P_2-K)\tilde{\Delta}(P_1 - K + Q_1).
\ee
Following the Eq.~(\ref{X1a2a}) it is possible here to show that
\be
Y_{i}^{1(a)}&=&-T\sum\limits_{n}\omega_n^i\tilde{\Delta}^2(K)\Delta(P_1 - K)\nn
&\times&\Delta(P_2-K)\Delta(P_1-K+Q_1)\nn
&=&\frac{\del X_i^{2(a)}}{\del m^2}.
\ee
With finite quark mass, $X_i^{2(a)}$'s can be calculated by generalizing the corresponding zero quark mass $X_i^{2(a)}$'s as
\be
X_0^{2(a)}&=&-\frac{1}{16k^3E_m}
\Bigg[\frac{n_B(E_m)+n_F^-(E_m)}{(P_1\mycdot\hat{K}')(P_2\mycdot\hat{K}')[
(P_1+Q_1)\mycdot\hat{K}']}\nn
&& \hspace{1cm}- 
\frac{n_B(E_m)+n_F^+(E_m)}{(P_1\mycdot\hat{K})(P_2\mycdot\hat{K})[
(P_1+Q_1)\mycdot\hat{K}]}\Bigg],
\label{X0_m}
\ee
\be
X_1^{2(a)}&=&\frac{i}{16k^2E_m}
\Bigg[\frac{n_B(E_m)+n_F^-(E_m)}{(P_1\mycdot\hat{K}')(P_2\mycdot\hat{K}')[
(P_1+Q_1)\mycdot\hat{K}']}\nn
&& \hspace{1cm} + 
\frac{n_B(E_m)+n_F^+(E_m)}{(P_1\mycdot\hat{K})(P_2\mycdot\hat{K})[(P_1 + 
Q_1)\mycdot\hat{K}]}\Bigg],
\label{X1_m}
\ee
\be
X_2^{2(a)}&=&\frac{1}{16kE_m}
\Bigg[\frac{n_B(E_m)+n_F^-(E_m)}{(P_1\mycdot\hat{K}')(P_2\mycdot\hat{K}')[
(P_1+Q_1)\mycdot\hat{K}']}\nn
&& \hspace{1cm} - 
\frac{n_B(E_m)+n_F^+(E_m)}{(P_1\mycdot\hat{K})(P_2\mycdot\hat{K})[(P_1+Q_1)\mycdot\hat{K}]}
\Bigg],
\label{X2_m}
\ee
\be
X_3^{2(a)}&=&-\frac{i}{16E_m}
\Bigg[\frac{n_B(E_m)+n_F^-(E_m)}{(P_1\mycdot\hat{K}')(P_2\mycdot\hat{K}')[
(P_1+Q_1)\mycdot\hat{K}']}\nn
&& \hspace{1cm} + 
\frac{n_B(E_m)+n_F^+(E_m)}{(P_1\mycdot\hat{K})(P_2\mycdot\hat{K})[(P_1+Q_1)\mycdot\hat
{K}]}\Bigg].
\label{X3_m}
\ee
In Eqs.~(\ref{X0_m}) -(\ref{X3_m}), $E_m$ represents $E_m=\sqrt{k^2+m^2}$. Note that we follow Ref.~\cite{Ayala:2014uua} and take the quark mass-dependent Bose-Einstein distribution function along with the Fermi-Dirac distribution function and the finite quark mass acts as an infrared regulator. Now, combining all the contributions 
from Eqs.~(\ref{X0_m})\,-\,(\ref{X3_m}), the tensor in Eq.~(\ref{tensor_def}) can be written as
\be
\delta\mathcal{T}_{\mu\nu\beta}^{1(a)} 
&=&-\frac{1}{32\pi^2}\left(\frac{\del}{\del 
y^2}\right)\int\limits_0^\infty 
\frac{x^2 dx}{\sqrt{x^2+y^2}}\bigg[2n_B(\sqrt{x^2+y^2})\nn
&&+\ n_F^+(\sqrt{x^2+y^2})+n_F^-(\sqrt{x^2+y^2})\bigg]\nn
&\times& \int\frac{d\Omega}{4\pi}\frac{K_\mu K_\nu K_\beta}
{(P_1\mycdot\hat{K})(P_2\mycdot\hat{K})[(P_1 + Q_1)\mycdot\hat{K}]},
\label{delta_tau}
\ee
where $k/T=x$ and $m/T=y$.

The integral over $x$ in Eq.~(\ref{delta_tau}) can be represented in terms of the functions
\be
f_n(y,\tilde{\mu})&=&\frac{1}{\Gamma(n)}\int\limits_0^\infty 
\frac{x^{n-1} dx}{\sqrt{x^2+y^2}} \frac{1}{2}\Bigg[\frac{1}{e^{\sqrt{x^2+y^2}-\tilde{\mu}}+1}\nn
&&+\frac{1}{e^{\sqrt{x^2+y^2}+\tilde{\mu}}+1}\Bigg],\\
h_n(y)&=&\frac{1}{\Gamma(n)}\int\limits_0^\infty 
\frac{x^{n-1} dx}{\sqrt{x^2+y^2}} \frac{1}{e^{\sqrt{x^2+y^2}}-1},
\ee
with $\tilde{\mu}=\mu/T=2\pi\hat\mu$.
So,
\be
\delta\mathcal{T}_{\mu\nu\beta}^{1(a)} &=& 
-\frac{1}{8\pi^2}\left(\frac{\del}{\del 
y^2}\right)\Big[f_3(y,\tilde{\mu})
+h_3(y)\Big]\nn
&&\times\int\frac{d\Omega}{4\pi}\frac{{\hat K}_\mu {\hat K}_\nu {\hat K}_\beta}
{(P_1\mycdot\hat{K})(P_2\mycdot\hat{K})[(P_1 + Q_1)\mycdot\hat{K}]}\nn
&=& \frac{1}{32\pi^2}\Big[f_1(y,\tilde{\mu})
+h_1(y)\Big]\nn
&&\times\int\frac{d\Omega}{4\pi}\frac{{\hat K}_\mu {\hat K}_\nu {\hat K}_\beta}
{(P_1\mycdot\hat{K})(P_2\mycdot\hat{K})[(P_1 + Q_1)\mycdot\hat{K}]},
\ee
where we have used the following identities:
\be
 \frac{\del h_{n+1}}{\del y^2}&=&-\frac{h_{n-1}}{2n},\nn
 \frac{\del f_{n+1}}{\del y^2}&=&-\frac{f_{n-1}}{2n}.
\label{diffeq}
\ee
In the HTL approximation ($m\ll T$) the two functions $f_1(y,\tilde{\mu})$ and $h_1(y)$ can be expanded as
\be
f_1(y,\tilde{\mu}) &=& -\frac{1}{2}\log\left(\frac{y}{4\pi}\right) + 
\frac{1}{4}\aleph(z) + \cdots ,\\
h_1(y) &=& 
\frac{\pi}{2y}+\frac{1}{2}\log\left(\frac{y}{4\pi}\right)+\frac{1}{2}\gamma_E + 
\cdots ,
\ee
and keeping the leading terms in $m$, we get
\be
\delta\mathcal{T}_{\mu\nu\beta}^{1(a)} &=& 
\frac{1}{32\pi^2}\left[-\frac{1}{4}\aleph(z)-\frac{\pi 
T}{2m}-\frac{\gamma_E}{2}\right]\nn
&\times&\int\frac{d\Omega}{4\pi}\frac{K_\mu K_\nu K_\beta}
{(P_1\mycdot\hat{K})(P_2\mycdot\hat{K})[(P_1 + Q_1)\mycdot\hat{K}]}.
\ee
Similarly, the tensor related to the second diagram $1(b)$ can be written as
\begin{widetext}
\be
\delta\mathcal{T}_{\mu\nu\beta}^{1(b)} &=& 
\frac{1}{32\pi^2}\left[-\frac{1}{4}\aleph(z)-\frac{\pi 
T}{2m}-\frac{\gamma_E}{2}\right]\int\frac{d\Omega}{4\pi}\frac{K_\mu K_\nu K_\beta}
{(P_1\mycdot\hat{K})(P_2\mycdot\hat{K})[(P_2 - Q_1)\mycdot\hat{K}]}.
\ee
Combining all the contributions, the magnetic field correction to the four-point QCD vertex can be written as
\be
\delta\Gamma_{\mu\nu}&=&4i\gamma_5g^2M(T,\mu,m,qB)\int\frac{d\Omega}{4\pi}
\left[\frac{1}{P_1\mycdot\hat{K}}+\frac{1}{P_2\mycdot\hat{K}}\right]\frac{1}
{[(P_1 + Q_1)\mycdot\hat{K}][(P_2-Q_1)\mycdot\hat{K}]}\nn
&\times&\Bigg[\Big\{(\hat{K}\mycdot b)(\hat{K}_\mu u_\nu + \hat{K}_\nu u_\mu) - 
(\hat{K}\mycdot u)(\hat{K}_\mu b_\nu + {\hat K}_\nu 
b_\mu)\Big\}\hat{\slashed{K}} 
+ \hat{K}_\mu \hat{K}_\nu\Big\{(\hat{K}\mycdot 
b)\slashed{u}-(\hat{K}\mycdot u)\slashed{b}\Big\}\Bigg].
\label{Gmn_final}
\ee
Note that the expression for $\delta\Gamma_{\mu\nu}$ in Eq.~(\ref{Gmn_final}) can be predicted using the expression of three-point quark-gluon vertex in Eq.~(\ref{Gmuexpl}) based on the Ward identity and symmetry property of $\delta\Gamma_{\mu\nu}$ with respect to the Dirac indices.

Now, we can check the Ward identity of the four-point quark-gluon vertex as
\be
Q_1^\nu\ 
\delta\Gamma_{\mu\nu}&=&4i\gamma_5g^2M(T,\mu,m,qB)\int\frac{d\Omega}{4\pi}
\left[\frac{1}{P_1\mycdot\hat{K}}+\frac{1}{P_2\mycdot\hat{K}}\right]\frac{1}
{[(P_1 + Q_1)\mycdot\hat{K}][(P_2-Q_1)\mycdot\hat{K}]}\nn
&\times&\Bigg[\Big\{(\hat{K}\mycdot b)(Q_1\mycdot u)-(\hat{K}\mycdot 
u)(Q_1\mycdot b)\Big\}\hat{K}_\mu\hat{\slashed{K}}
+\Big\{(\hat{K}\mycdot b) u_\mu-(\hat{K}\mycdot u) 
b_\mu)\Big\}\hat{\slashed{K}}Q_1\mycdot\hat{K}\nn
&&+ \hat{K}_\mu \Big\{(\hat{K}\mycdot b)\slashed{u}-(\hat{K}\mycdot 
u)\slashed{b}\Big\}Q_1\mycdot\hat{K}\Bigg].
\label{WI_34a}
\ee
Now, the first term of the vector structure of Eq.~(\ref{WI_34a}) can be neglected in the spirit of the HTL approximation. The remaining terms can be decomposed as
\be
Q_1^\nu\ \delta\Gamma_{\mu\nu}&\simeq& 
4i\gamma_5g^2M(T,\mu,m,qB)\int\frac{d\Omega}{4\pi}
\left[\frac{1}{(P_1\mycdot\hat{K})[(P_2-Q_1)\mycdot\hat{K}]} - 
\frac{1}{(P_2\mycdot\hat{K})[(P_1+Q_1)\mycdot\hat{K}]}\right]\nn
&&\times\left\{\left[(\hat{K}\mycdot b)\hat{\slashed{K}}u_\mu - 
(\hat{K}\mycdot u)\hat{\slashed{K}}b_\mu\right] + \left[(\hat{K}\mycdot b)\slashed{u}
- (\hat{K}\mycdot u)\slashed{b}\right]\hat{K}_\mu\right\}\nn
&=&\delta\Gamma_\mu(P_1,P_2-Q_1)-\delta\Gamma_\mu(P_2,P_1+Q_1),
\ee
\end{widetext}
which clearly satisfies the Ward identity.
\section{Conclusion and outlook}
\label{sec:conclusion}
In this article, we have calculated the weak magnetic field correction to the four-point quark-gluon vertex at finite temperature and chemical potential. Additionally, we have recalculated the quark self-energy and also the three-point quark-gluon vertex at finite temperature and chemical potential in the presence of weak magnetic field. We have explicitly shown that only QED-like diagrams [1(a) and 1(b)] contribute to the four-point function, whereas the overall contribution from the remaining diagrams (pure QCD diagrams) [2(a), 2(b), 3(a), 3(b), and 4] is zero. The resultant four-point vertex satisfies the Ward identity with the three-point vertex. Also, the expressions presented herein for two- and three-point vertices satisfy the Ward identity within themselves.

These $N$-point functions will be necessary to calculate various quantities such as thermodynamics of the hot and dense medium in the presence of magnetic field. Looking to the future, we are planning to compute NNLO thermodynamics within HTLpt in the presence of weak magnetic field.
\section{Acknowledgments}
\label{sec:ack}
The author gratefully acknowledges the financial support from Alexander von Humboldt  Foundation, Germany, for the postdoctoral fellowship. The author also acknowledges useful discussions with A.~Bandyopadhyay, M.~G.~Mustafa, C.~A.~Islam, S.~Seth, and S.~Chakraborty.

\end{document}